\documentclass[aps,prl,twocolumn,showpacs,amsmath,superscriptaddress,groupedaddress]{revtex4-1}
\usepackage{graphics}
\usepackage{epsfig}
\usepackage[colorlinks=true,citecolor=blue,linkcolor=red,linktocpage=true,pagebackref=false]{hyperref}
\usepackage{color,soul} 

\begin{document}
   \title{
Long-range Fermi sea correlations as the resource for encoding quantum information
         }
\author{Michael Moskalets}
\email{michael.moskalets@gmail.com}
\affiliation{Department of Metal and Semiconductor Physics, NTU ``Kharkiv Polytechnic Institute", 61002 Kharkiv, Ukraine}

\date\today
\begin{abstract}
An extra constant phase can be added to the wave function of a finite-length portion of a chiral Fermi sea. 
This phase can be read-out with the help of an imbalanced  interferometer, where such a phase carrier interferes with the reference Fermi sea. 
As a result of such interference, the same in value but opposite in sign charge is appeared at interferometer's outputs. 
A phase carrier consists of electron-hole pairs residing on the surface of the Fermi sea. 
Importantly, these pairs are not only electrically neutral, but in addition do not carry heat.  
A phase carrier can be created, for instance, with the help of an on-demand  single-electron source able to produce excitations with a multiple-peak density profile.  
\end{abstract}
\maketitle

The potential to shape a single-electron wave-function, which was discussed in Ref.~\onlinecite{Ott:2014vs}, extends to a new dimension the realm of quantum coherent electronics, a new and exciting branch of mesoscopics, whose history started with implementation of on-demand single-electron sources\cite{Blumenthal:2007ho,Feve:2007jx,Fujiwara:2008gt,Kaestner:2008gv,Dubois:2013dv}. 
The experimental demonstration of quantized charge pumping\cite{Wright:2008kc,Kaestner:2008cm,Giblin:2012cl,Fricke:2013we}, of a quantum optics like  behavior\cite{{Bocquillon:2012if},Bocquillon:2013dp,Dubois:2013dv,Fletcher:2013kt,{Bocquillon:2013fp}}  (see also theoretical discussions in Refs.~\onlinecite{Olkhovskaya:2008en,{Splettstoesser:2010bo},{Grenier:2011js},Haack:2011em,{Grenier:2011dv},{Jonckheere:2012cu},Haack:2013ch,Dubois:2013fs,Ferraro:2013bt}), and very recently of partitioning of generated on-demand pairs of electrons\cite{Ubbelohde:2014vx} makes quantum coherent electronics a promising platform for both quantum metrological\cite{Pekola:2013fg} and quantum information applications\cite{Bennett:2000kl}. 

A time-bin qubit is promising for encoding quantum information.\cite{Franson:1989uu,{Brendel:1999uh},Humphreys:2013eh,Donohue:2013da}  
A single particle having a multiple-peak density profile seems to be the most compact  time-bin qubit.  
The possibility to encode quantum information into the properly shaped single photons was already demonstrated.\cite{Vasilev:2010ca,{NisbetJones:2013ks}} 
With single electrons in solid state systems this is still to be done. 
The essential requirement is a necessity to handle an electron wave function phase. 
The approach utilizing an on-fly change of an electron phase was demonstrated in Ref.~\onlinecite{Yamamoto:2012bp} but with a limited visibility. 
Another approach is to impose a desired phase during creation of a single-electron  time-bin qubit.\cite{Ott:2014vs} 

In this Letter I describe a specific only for fermionic systems way to add a phase to a time-bin qubit. 
Namely, the phase is added not to a particle itself but to a portion of the Fermi sea moving together with a single electron having a multiple-peak profile. 

Before presenting an example let us clarify the effect of a constant phase added to the portion of the Fermi sea. 
For this purpose let us consider the first-order correlation function for electrons in a one-dimensional (1D) wave-guide, $\tilde {\cal G}^{(1)}\left(1,2  \right) =
\langle \hat\Psi^{\dag}(1) \hat\Psi(2) \rangle$, where $\hat\Psi (j) \equiv \hat\Psi\left(x_{j}t_{j} \right)$ is an electron field operator in second quantization evaluated at point $x_{j}$ and time  $t_{j}$,  $j=1,2$. 
The quantum-statistical average $\langle \dots \rangle$ is taken over the equilibrium state of the reservoir with the Fermi energy $ \mu$ the wave-guide is attached to. 
At zero temperature the correlation function reads, $\tilde {\cal G}^{(1)} _{ \mu} \left(1,2  \right) = \exp \left(  i [ \phi _{2} - \phi _{1}  ]  \right)  {\cal G}^{(1)} _{ \mu} \left(1,2  \right)$ (see e.g. Ref.~\onlinecite{Grenier:2011js}), with  $\phi_{j} = -\mu t_{j}/\hbar + k _{ \mu} x_{j}$ and
\begin{eqnarray}
{\cal G}^{(1)} _{ \mu}\left(1,2  \right) 
= 
\frac{i  }{2 \pi v _{ \mu} } \frac{ 1 }{ \tau _{2} - \tau _{1}  + i 0 ^{+}  } .
\label{Gmu}
\end{eqnarray}
Here $ \tau _{j} = t _{j} - x _{j} / v _{ \mu}$;  $v _{ \mu}$ and $k _{ \mu}$ the velocity and the wave number evaluated at the Fermi energy $ \mu$. 
Let us create a phase carrier (PhC) by adding a constant phase $\chi$ to a finite portion of length $\ell$ of a chiral Fermi sea.
If we do it adiabatically such that the equilibrium is not destroyed (an example is discussed later on in this Letter), then the correlation function becomes, 
\begin{eqnarray}
{\cal G}^{(1)} _{ \chi}\left(1,2  \right) 
=  
e ^{i \Xi(1,2) } {\cal G}^{(1)} _{ \mu}\left(1,2  \right) .
\label{Gchi}
\end{eqnarray}
Here the subscript $ \chi$ indicates the added phase; 
The phase $\Xi(1,2) =  \chi$ if $x _{2}$ belongs to the segment $\ell$ while $x _{1}$ does not, $\Xi(1,2) =  -\chi$ if $x _{1}$ belongs to $\ell$ while $x _{2}$ does not, and  $\Xi(1,2) = 0$ otherwise. 
Note that the PhC propagates with velocity $v _{ \mu}$. 

\begin{figure}[b]
\includegraphics[width=80mm]{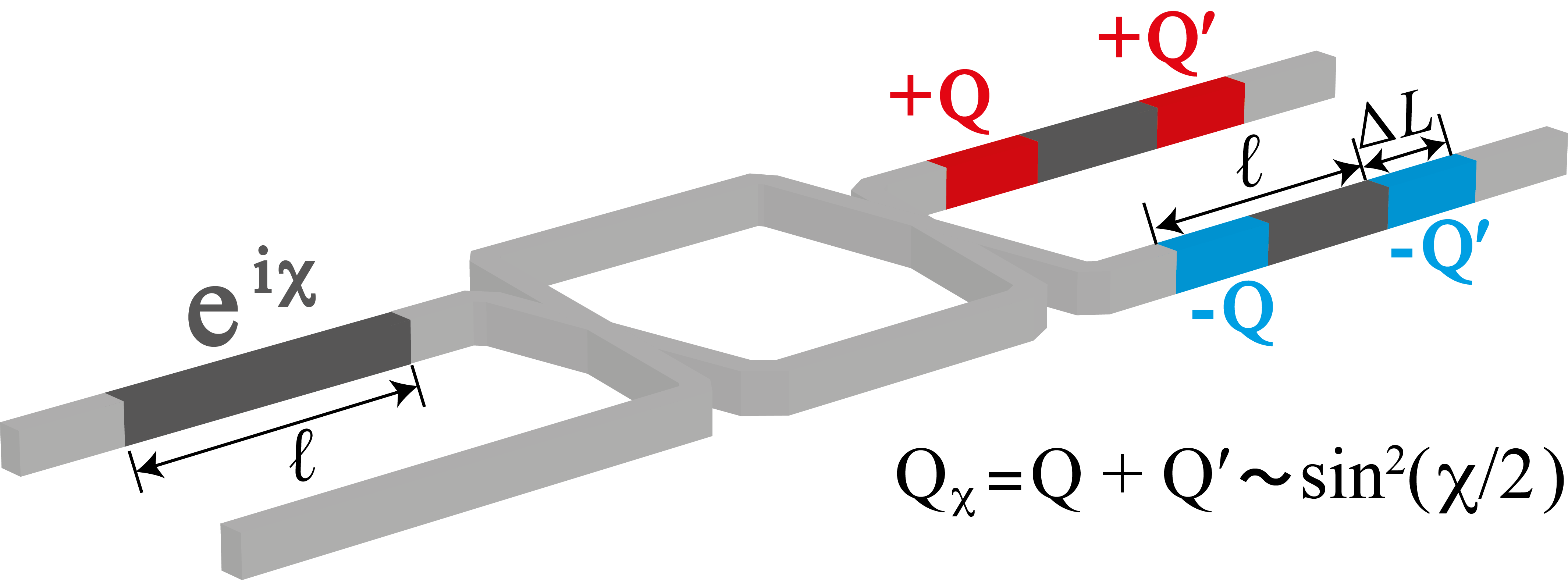}
\caption{(Color online) A sketch of a Mach-Zehnder interferometer with the arm difference $\Delta L > 0$ made of 1D chiral electronic wave-guides (light grey). When a portion of the Fermi sea of a finite length $\ell > \Delta L$ (dark grey) having an extra phase $\chi$ passes through the interferometer, an opposite charge $Q _{ \chi} =  Q + Q ^{\prime}$ (carried by segments of length $\Delta L$ shown in red and blue) is transferred to output leads.}
\label{fig1}
\end{figure}

If now we let such an adiabatic PhC to pass a Mach-Zehnder interferometer, then we find that an opposite charge $Q_{\chi}$ is transferred to the two output leads as shown in Fig.~\ref{fig1}. 
To calculate it we express the interference current $I ^{int}(t)$ (evaluated at some fixed point $x$ behind the interferometer) in terms of the first-order correlation function (see e.g. Refs.~\onlinecite{Haack:2011em,Haack:2013ch}), 
\begin{eqnarray}
I ^{int}(t) = 
2 \gamma e \, {\rm Re} \left\{ e ^{-i \Delta \phi} \Delta {\cal G}^{(1)} \left(t - \tau _{up}, t - \tau _{down}  \right) \right\} .
\label{Iint}
\end{eqnarray}
Here $\Delta {\cal G}^{(1)}(t _{1}, t _{2})  = {\cal G}^{(1)} _{ \chi}(t _{1}, t _{2}) - {\cal G}^{(1)} _{ \mu}(t _{1}, t _{2}) $ for $x _{1} = x _{2} = x$;
$ \gamma = \sqrt {R _{L} T _{L}R _{R} T _{R} }$ with $T _{ \alpha} = 1 - R _{ \alpha}$ being the transmission of the wave-splitter $ \alpha = L, R$ forming the interferometer; 
$ \Delta \phi = 2 \pi \Phi/\Phi _{0} + k _{ \mu} \Delta L$ is a phase difference acquired by an electron with energy $ \mu$ traversing the interferometer along  the lower (with length $L _{down}  = v _{ \mu} \tau _{down} $) and the upper (with length $L _{up} = L _{down}  + \Delta L = v _{ \mu} \tau _{up}$) arms of the interferometer; 
$\Phi$ is a magnetic flux threading the interferometer; 
$\Phi _{0} = h/e$ is the magnetic flux quantum. 
After integrating over time (longer than $(\Delta L + \ell)/v _{ \mu}$) we find a transferred charge, $Q _{ \chi} =  \int dt I ^{int}(t)   $, to be the following, 
\begin{eqnarray}
\frac{Q _{ \chi}  }{ e }
&=& 
\xi \frac{4 \gamma }{ \pi } \sin ^{2} \left( \frac{ \chi }{2 } \right)  \sin\left( \Delta \phi  \right) \,, 
\label{Qchi} \\
{\rm with} 
\nonumber \\
\xi 
&=& 
\left\{
\begin{array}{ll}
1 \,,  & \Delta L < \ell \,,   \\
\ \\
 \frac{\ell }{\Delta L } \,, &   \Delta L > \ell \,.
\end{array}
\right.
\nonumber
\end{eqnarray}
For $ 0 < \Delta L < \ell$ the factor $ \xi$ is constant. 
This is due to the cancellation of two factors $\Delta L$, one is in the denominator and one is in the numerator of the equation for $Q _{ \chi}$. 
The factor $\Delta L$ in the numerator arises since the charge $Q _{ \chi}$ is proportional to the length over which the PhC interferes with the reference Fermi sea. 
The factor  $\Delta L$ in the denominator arises from the Fermi sea propagator, Eq.~(\ref{Gmu}), where we should use $ \tau _{2} - \tau _{1} = \Delta L/v _{ \mu}$.  
Therefore, for $\Delta L < \ell$ the charge $Q _{ \chi}$ depends on the phase $ \chi$ but not on the length $\ell$ of a PhC. 
This fact makes it possible to read out the phase encoded in the Fermi sea for a wide range of length $\ell$.

The charge $Q _{ \chi}$ can be thought as arising due to interferometric filtering of electron-hole pairs created when the phase $ \chi$ was added.  
The number of such pairs, $N _{ep}$, can be estimated from the maximum transferred charge as follows, $N _{ep} \geq \sin ^{2} \left( \chi / 2 \right) /\pi$, where we use in Eq.~(\ref{Qchi}) the value $4 \gamma = 1$ for an interferometer with the maximum filtering efficiency, $T _{L} = T _{R} = 0.5$, and put $\sin(\Delta\phi) = 1$. 

{\it Creation of a PhC.---}
Now I give an example of how an adiabatic phase carrier can be created. 
This method is essentially based on the observation made in Refs.~\onlinecite{Juergens:2011gu,Rossello} that the single-electron source changes the phase of electrons passing by it in the wave-guide this source is attached to. 
To create an adiabatic PhC the single-electron source of Ref.~\onlinecite{Feve:2007jx} working in the adiabatic regime and driven by the properly designed gate voltage can be used. 
The source is a quantum capacitor\cite{Buttiker:1993wh,{Gabelli:2006eg}} made of a circular edge state of a two-dimensional electron gas in the integer quantum Hall effect regime. 
The capacitor is connected via a quantum point contact (QPC) with transmission $T_{QPC}$ to a nearby linear edge state playing the role of an electron wave-guide. 
The periodic voltage $U(t) = U(t + {\cal T})$ applied to a top gate drives quantum levels of a capacitor up and down (in energy) such that only one level crosses the Fermi level of a wave-guide during the period ${\cal T}$. 
When the capacitor's level rises above the Fermi level, an electron leaves the capacitor. 
While when the capacitor's level sinks below the Fermi level, an electron enters the capacitor and, correspondingly, a hole is left in the wave-guide. 
Therefore, a periodically driven capacitor generates a quantized ac current consisting of the stream of alternating single electrons and holes. 

Due to the coupling to a wave-guide, the quantum levels of a capacitor are widened. 
If the voltage $U(t)$ changes so slow that the crossing time  (the time during which a widened level crosses the Fermi level) is large compared to the dwell time (the time during which an electron escapes a capacitor, $ \tau _{D} = h/(T _{QPC} \mathit{\Delta})$, where $\mathit{\Delta}$ is the level spacing), then such a regime is called adiabatic.\cite{Splettstoesser:2008gc,Moskalets:2013dl}  
In this regime at zero temperature the generated current, $I(t) = e(-i/2\pi) S \partial S ^{*}/ \partial t$\cite{Buttiker:1994vl,Avron:2000de,Buttiker:2006wt}, is expressed in terms of the scattering amplitude $S(t)$ for electrons with energy $ \mu$ propagating in the wave-guide and passing by the capacitor. 
The amplitude $S$ depends on time since it depends on the position of quantum levels of the capacitor. 
In the single-channel chiral case under consideration the unitarity requires $\left | S(t) \right | ^{2} =1$ hence $S(t) = \exp[i \chi(t)]$. 
Therefore, the effect of $S(t)$ is only a change of the phase of the wave function of electrons passing by the capacitor.\cite{Moskalets:2014ea}

\begin{figure}[t]
\includegraphics[width=80mm]{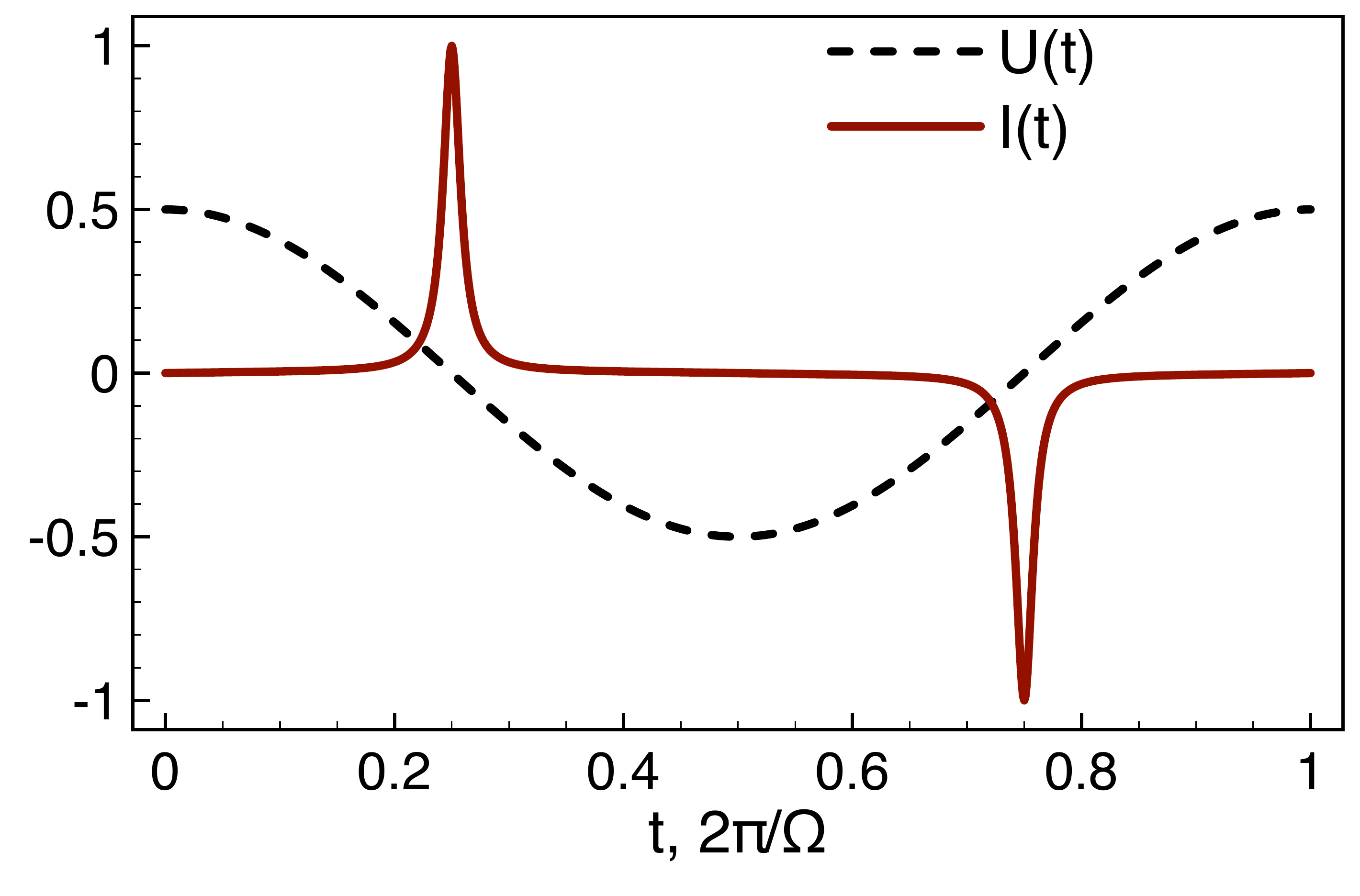}
\includegraphics[width=80mm]{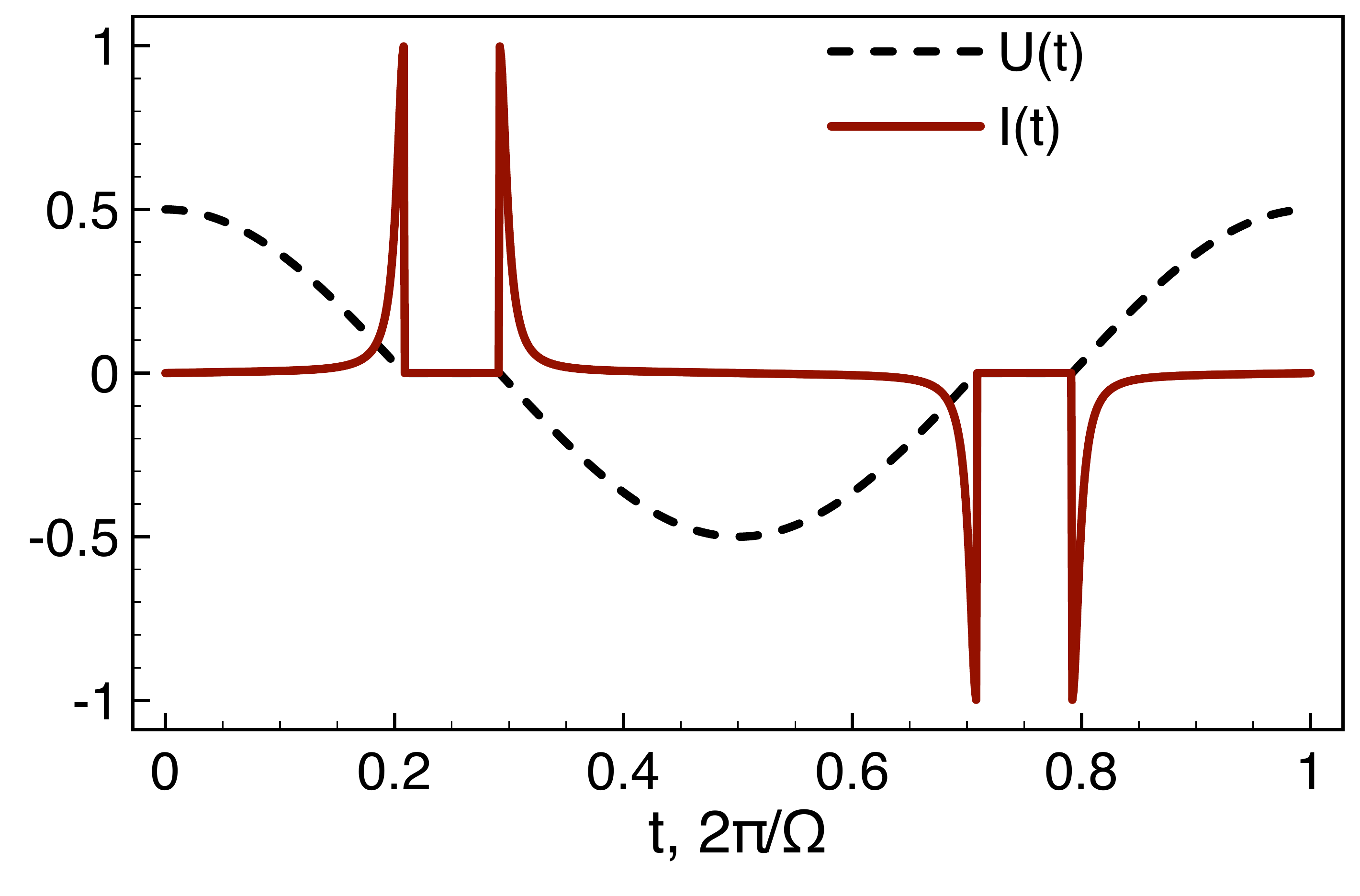}
\caption{A time-dependent current $I(t)$ generated by a single-electron source driven by the voltage $U(t)$. Time $t$ is given in units of the period ${\cal T} = 2\pi/\Omega$. When the quantum level of the source crosses the Fermi level of the electronic wave-guide, an electron ($t _{e} = 0.25$) and a hole ($t _{h} = 0.75$) are emitted. {\bf Upper panel:} When the voltage is varied continuously, the emitted particle has a density profile with one peak. {\bf Lower panel:} When the voltage is frozen for a while at the time of crossing, the emitted particle has a density profile with two peaks  separated by the flat part of duration $\tau _{delay} = 0.1 {\cal T}$ in the present case. Importantly this flat part has an extra phase $\pi$ compared to the Fermi sea residing outside the particle.}
\label{fig2}
\end{figure}

If the driving potential changes continuously, say, $U(t) = U \cos (\Omega t)$, $\Omega = 2\pi /{\cal T}$, then the emitted particles (electrons and holes) have a  Lorentzian density profile and the corresponding current pulses (of opposite sign for electrons and holes) are Lorentzian as shown in the upper panel of Fig.~\ref{fig2}.\cite{Keeling:2006hq,Keeling:2008ft} 
In contrast, if we would stop the driving potential $U(t)$ at the time when the capacitor's level crosses the Fermi level and after the time delay $\tau_{delay}$ would continue to change it, then the emitted particle would have a double-peak density profile, see Fig.~\ref{fig2}, the lower panel.

In the limit of a small transmission, $T_{QPC} \ll 1$, the scattering amplitude close to the time of an electron emission, $t = t _{e}$, is the following,
\begin{eqnarray}
S(t) = 
\left\{
\begin{array}{ll}
\dfrac{t - t _{e} + i \Gamma}{t - t _{e} - i \Gamma} \,, & t - t _{e} < 0 \,, \\
\ \\
e ^{i \pi} \,, & 0< t - t _{e} < \tau_{delay} \,, \\
\ \\
\dfrac{t - t _{e} - \tau_{delay} + i \Gamma}{t - t _{e} - \tau_{delay} - i \Gamma} \,, & \tau_{delay} < t - t _{e}\,.
\end{array}
\right.
\label{st}
\end{eqnarray}
Here the parameter $\Gamma$ characterizes the width of an emitted single-particle state.
If the amplitude of the driving potential is equal to half of the level spacing in the capacitor and the level oscillates symmetrically with respect to $ \mu$, then $\Gamma = T_{QPC}/(4\pi^2)$. 
Close to the time of a hole emission, $t = t_{h}$, the following scattering amplitude should be used, $S_{h}(t) = S^{*}(t + t _{e} - t_{h})$. 
The corresponding time-dependent current (at some fixed position behind the source) for one period is shown  in  Fig.~\ref{fig2} for $\tau _{delay} = 0$ (the upper  panel) and $\tau _{delay} \ne 0$ (the lower panel). 
Importantly, at $\tau _{delay} \ne 0$ the current pulse corresponding to a single particle has a double-peak profile. 
As it is seen from Eq.~(\ref{st}), the flat inner part of length $\ell = v _{ \mu} \tau _{delay}$ has an extra phase $ \chi =  \pi$ (since $S(t) = \exp(i \pi)$ for $0< t - t _{e} < \tau_{delay}$), compared to what is far away from the place where the emitted particle propagates (since $S(t) = 1$ for $(\tau_{delay} + \Gamma) \ll |t - t _{e}|$). 
Therefore, this flat part is an adiabatic PhC in question.  

From the surrounding Fermi sea the PhC is separated by semi-pulses corresponding to an emitted particle. 
Importantly, for electrons emitted adiabatically, the excess energy (i.e., the energy counted from $ \mu$ or,  simply, heat) is proportional to the squared electrical current integrated over time, that is nothing but the well-known Joule-Lentz law.\cite{Moskalets:2009dk,Moskalets:2013dl}  
Therefore, heat carried by either a single-peak particle (upper panel in Fig.~\ref{fig2}) or a two-peak particle (lower panel in Fig.~\ref{fig2}) is the same, $\hbar/(2\Gamma)$. 
There is no heat associated with a flat inner part hence with a PhC. 
We conclude, electron-hole pairs constituting the PhC have no excess energy, i.e. they reside at the surface of the Fermi sea. In view of this, I call them \emph{zero-energy electron-hole pairs}.

{\it Characterization of a PhC.---}
To demonstrate how the phase, $ \chi = \pi$, of such a PhC can be read out, let us use an electronic Mach-Zehnder interferometer\cite{Ji:2003ck,Litvin:2007kr,Roulleau:2007ks,Bieri:2009ki} with the source being placed at one of its input leads.  
The current at one of the output leads, $I ^{out}(t)  = I ^{cl}(t) + I ^{int}(t)$, is the sum of the classical part, $ I ^{cl}(t) = R _{L} R _{R} I(t - \tau _{up}) + T _{L} T _{R} I(t - \tau _{down} )$, and the interference part, $I ^{int}(t)$, given in Eq.~(\ref{Iint}) with $ \Delta {\cal G}^{(1)}$ expressed in terms of the scattering amplitude as follows:\cite{Haack:2011em,Haack:2013ch}
\begin{eqnarray}
 \Delta {\cal G}^{(1)}\left( t _{1}, t _{2} \right) 
&=& 
\frac{i e ^{ i ( \phi _{2} - \phi _{1}  )}  }{2 \pi v _{ \mu} } \frac{ S ^{*}(t _{1}) S(t _{2}) - 1 }{ t _{2} - t _{1}   } .
\label{dg}
\end{eqnarray}
The current at another output lead is $R _{L} T _{R} I(t - \tau _{up})  + T _{L} R _{R} I(t - \tau _{down} ) - I ^{int}(t)$. 

\begin{figure}[t]
\includegraphics[width=80mm]{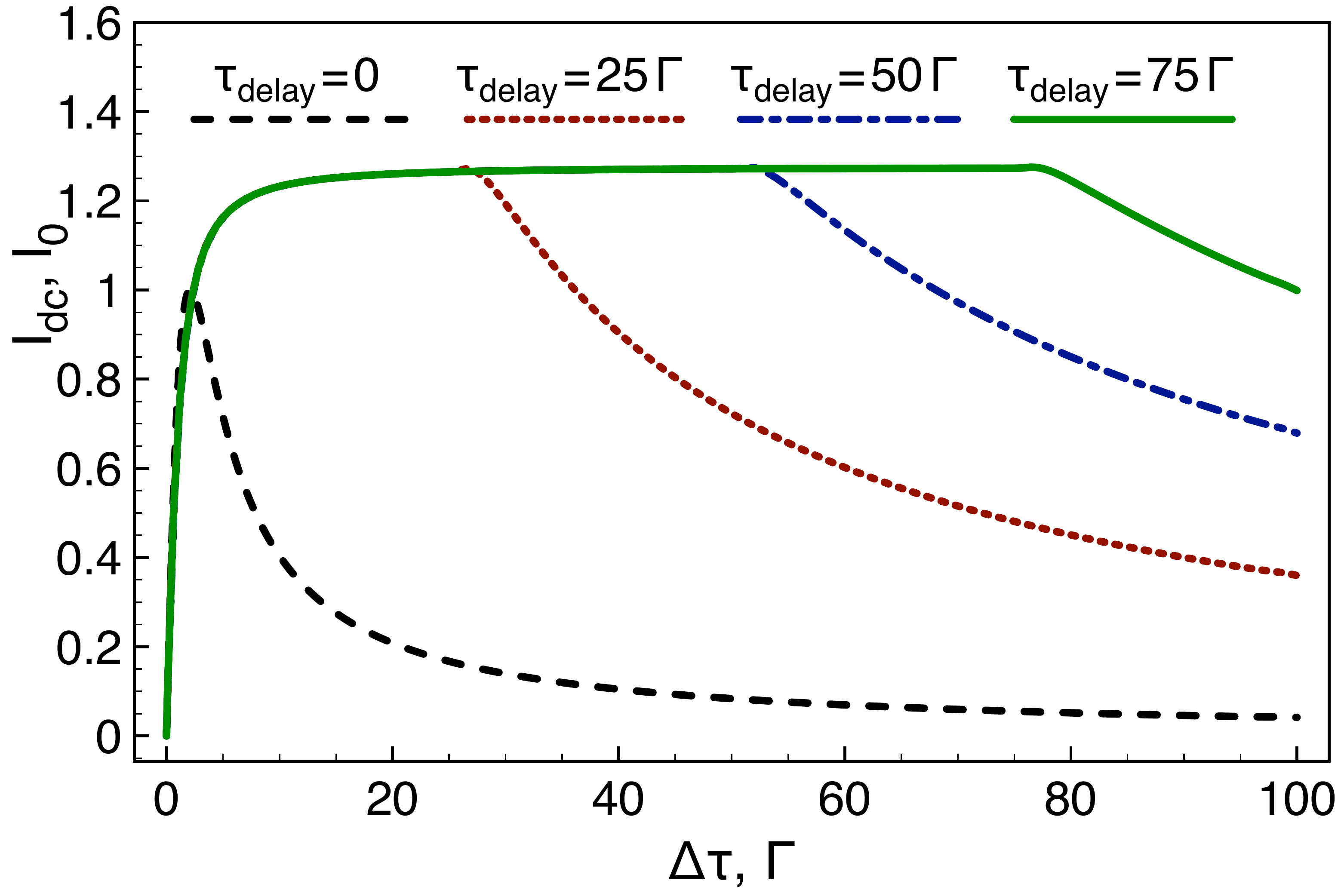}
\caption{The maximum value of a dc current, $I_{dc}$, measured at one of the interferometer's output is shown as a function of the imbalance $\Delta\tau = \Delta L / v _{\mu}$ for several values of the duration $\tau _{delay}$ of the sleep stage of a driving potential. The plateau due to zero-energy electron-hole pairs lasts until $\Delta \tau < \tau _{delay}$. $\Gamma$ is the half width of a single-electron excitation.  $I _{0} = 2 \gamma e/ {\cal T}$ with ${\cal T}$  the period of the driving potential and the parameter $\gamma$ defined after Eq.~(\ref{Iint}).}
\label{fig3}
\end{figure}

To suppress the contribution due to current pulses and to get only the contribution due to the PhC, the dc current at the output contact should be measured, $I _{dc} = \int _{0} ^{{\cal T}} I ^{int}(t) dt/{\cal T}$. 
The classical part of a current has no a dc part, since the source emits the same number of electrons and holes whose contributions cancel each other. 
The interference part of a dc current consists of two physically different contributions. 
One is due to single particles (electrons and holes) emitted by the single-electron  source. 
This contribution exists if the current pulse overlaps with itself after passing the interferometer. 
Therefore, it drops quickly down when the interferometer imbalance $\Delta L$ becomes larger than the width $\Gamma$ of a current peak (see the black dashed  line in Fig.~\ref{fig3}).  
The other contribution is due to zero-energy electron-hole pairs belonging to a PhC created during the sleep stage of duration $ \tau _{delay} > 0$, when the driving potential is frozen. 
Such a contribution exists if the PhC does not fully overlap with itself after passing the interferometer. 
It results in the universal (i.e., independent of $\Delta L = v _{ \mu} \Delta \tau$ and $\ell = v _{ \mu} \tau _{delay}$) dc current once the following inequality holds: $\Gamma < \Delta\tau < \tau _{delay}$. 

In Fig.~\ref{fig3} the maximum value of $I _{dc}$ is shown as a function of $\Delta \tau$. 
This maximum value is nothing but the amplitude of oscillations of a dc current as a function of a magnetic flux threading the interferometer with all other parameters being fixed. 
The value of a current at the plateau, $I _{dc} = I _{0}4/ \pi = 2 Q _{ \chi}/ {\cal T}$ with $I _{0} = 2 \gamma e/ {\cal T}$, is due to a contribution of two PhCs (hence the factor $2$) produced during each period, one associated to an electron and the other one is associated to a hole.
This value is in full agreement with Eq.~(\ref{Qchi}) defining $Q _{ \chi}$.  
Notice that in the case under consideration $ \chi = \pi$. 

To vary the phase $ \chi$ carried by the PhC, the beginning of the sleep stage should be varied. 
If during an electron emission stage the potential becomes frozen not at $t = t _{e}$ but at $t = t _{e} + t _{0}$, then, according to Eq.~(\ref{st}), the phase of the PhC is changed from $ \chi =  \pi$ to $ \chi = -i \ln([ t _{0} + i \Gamma]/[t _{0} - i \Gamma]) = \arctan(2\Gamma t _{0}/[ t _{0} ^{2} - \Gamma ^{2} ] )$. 
In this case the value of a current at the plateau in Fig.~\ref{fig3} becomes $I _{dc} = I _{0} \left[  (4/\pi)\sin ^{2}( \chi/2) + (4\Gamma/\Delta\tau) \cos ^{2}( \chi/2) \right]$, which in the limit of $\Delta\tau \gg \Gamma$ agrees with Eq.~(\ref{Qchi}).
The term proportional to $\cos ^{2}( \chi/2)$ is due to the remaining contribution of single-particle semi-pulses. 

The method of creation and characterization of a PhC proposed here seems to be feasible with current-day technology. 
For the electronic circuits built with edge states of the quantum Hall regime the main limiting factor is the decoherence length.\cite{Ferraro:2014ur} 
At reasonable low temperatures of order few tens mK this length varies from $20\, \mu$m\cite{Roulleau:2008gp} up to $80\, \mu$m\cite{Altimiras:2010dk}. 
Therefore, one can choose the interferometer imbalance $\Delta L \sim 10\, \mu$m.
Using the typical velocity of excitations in edge states $v _{ \mu} = 10 ^{5}\,$m/s\cite{Sukhodub:2004bh,Gabelli:2007ev}, one can find $\Delta\tau = \Delta L/ v _{ \mu} \sim 10 ^{-10}\,$s. 
The dwell time of a single-electron source can be made as small as $\tau _{D} =20\times 10^{-12}\,$s.\cite{Mahe:2010cp}
For the single-particle emission to be adiabatic it is required that $\Gamma > \tau _{D}$. 
Therefore, it is $\Delta\tau/\Gamma < 5$. 
This is almost at the limit necessary to observe a plateau in Fig.~\ref{fig3}. 
To increase this ratio  a quantum dot with a larger level spacing should be used as an electron source. 
On the other hand, to improve the visibility of the contribution due to zero-energy electron-hole pairs (the plateau in Fig.~\ref{fig3}), the squared current can be measured. 
In this case the contribution due to single particles drops down at smaller values of the ratio  $\Delta\tau/\Gamma$, see the discussion in Ref.~\onlinecite{Haack:2011em}. 

Finite temperatures destroy long-range correlations in the Fermi sea. 
However, if the thermal length $\lambda _{th} = \hbar v _{\mu}/k _{B}\theta$ is larger than $\Delta L$, then Eq.~\ref{Gchi} can still be used and the discussed effect should be observable. 
For $\Delta L \sim 10\, \mu$m this requires the temperature to be less than $\theta < 70\,$mK, which is compatible with the working temperature of a single-electron source.\cite{Parmentier:2012ed}

In conclusion, I proposed a method to change locally the phase of the wave function of the Fermi sea. 
The section of the Fermi sea of length $\ell$ with an extra phase $\chi$ plays the role of a phase carrier, which can transfer phase information to a distant place, where it can be read out with the help of an interferometer with a finite imbalance $\Delta L > 0$. 
Such an interferometer plays the twofold role.  
On one hand, it filters out the contribution of endpoints, where the Fermi sea's extra  phase changes from zero to $\chi$ and back, that results in time-dependent electrical currents. 
On the other hand, the interferometer splits neutral electron-hole pairs constituting a phase carrier and residing on the surface of the Fermi sea within the section $\ell$. 
The number of these pairs and, correspondingly, the generated dc current, depends on $\chi$ but neither on the length of a phase carrier $\ell$  nor on the interferometer's imbalance $\Delta L$. 
A Fermi sea based phase carrier could serve as a universal tool for transporting quantum information in coherent electronic circuits.

\begin{acknowledgments}
I am thankful to J. R. Ott, G. Haack, and J. Splettstoesser for numerous discussions and comments on the manuscript. 
\end{acknowledgments}

\end{document}